# An Average Analysis of Backtracking on Random Constraint Satisfaction Problems[*]

Ke Xu and Wei Li

*National Laboratory of Software Development Environment*
*Department of Computer Science and Engineering*
*Beijing University of Aeronautics and Astronautics, Beijing 100083, P.R. China*

Email[#]: {kexu,liwei}@nlsde.buaa.edu.cn

**Abstract.** In this paper we propose a random CSP model, called Model GB, which is a natural generalization of standard Model B. It is proved that Model GB in which each constraint is easy to satisfy exhibits non-trivial behaviour (not trivially satisfiable or unsatisfiable) as the number of variables approaches infinity. A detailed analysis to obtain an asymptotic estimate (good to $1+o(1)$) of the average number of nodes in a search tree used by the backtracking algorithm on Model GB is also presented. It is shown that the average number of nodes required for finding all solutions or proving that no solution exists grows exponentially with the number of variables. So this model might be an interesting distribution for studying the nature of hard instances and evaluating the performance of CSP algorithms. In addition, we further investigate the behaviour of the average number of nodes as $r$ (the ratio of constraints to variables) varies. The results indicate that as $r$ increases, random CSP instances get easier and easier to solve, and the base for the average number of nodes that is exponential in $n$ tends to 1 as $r$ approaches infinity. Therefore, although the average number of nodes used by the backtracking algorithm on random CSP is exponential, many CSP instances will be very easy to solve when $r$ is sufficiently large.

---

[*] Research supported by National 973 Project of China Grant No.G1999032701.

[#] For the first author, there is a permanent email address that is ke.xu@263.net.

# 1. Introduction

A *constraint satisfaction problem* (CSP) consists of a finite set $U = \{u_1, \cdots, u_n\}$ of $n$ variables and a set of constraints. For each variable $u_i$ a *domain* $D_i$ with $d_i$ elements is specified; a variable can only be assigned a value from its domain. For $2 \leq k \leq n$ a *constraint* $C_{i1,i2,\cdots,ik}$ consists of a subset $\{u_{i1}, u_{i2}, \cdots, u_{ik}\}$ of $U$ and a relation $R_{i1,i2,\cdots,ik} \subseteq D_{i1} \times \cdots \times D_{ik}$, where $i1, i2, \cdots, ik$ are distinct. $C_{i1,i2,\cdots,ik}$ is called a $k$-*ary* constraint which bounds the variables $u_{i1}, \cdots, u_{ik}$. $R_{i1,i2,\cdots,ik}$ specifies all the allowed tuples of values for the variables $u_{i1}, \cdots, u_{ik}$ which are compatible with each other. A *solution* to a CSP is an assignment of a value to each variable from its domain such that all the constraints are satisfied. A constraint $C_{i1,i2,\cdots,ik}$ is satisfied if the tuple of values assigned to the variables $u_{i1}, \cdots, u_{ik}$ is in the relation $R_{i1,i2,\cdots,ik}$. A CSP that has a solution is called *satisfiable*; otherwise it is *unsatisfiable*. Sometimes, it is desired to determine whether a CSP is satisfiable. However, in this paper we focus on the task of finding all solutions or proving that no solution exists.

CSP has not only important theoretical value in Artificial Intelligence, but also many immediate applications in areas ranging from vision, language comprehension to scheduling and diagnosis[6]. In general, CSP tasks are computationally intractable (NP-hard).

A simple algorithm for solving a CSP is backtracking. Backtracking works with an initially empty set of compatible instantiated variables and tries to extend the set to a new variable and a value for the variable. The most basic form of backtracking[21] analyzed in this paper is as follows:

*Algorithm: Backtracking*

*Input: A random CSP instance $F$*

*Output: All solutions to the instance $F$*

*1. Set $i \leftarrow 0$.*

*2. If $F(u_1/a_1, \cdots, u_l/a_l, \cdots, u_i/a_i)$ is false, go to 6. ($F(u_l/a_l)$ stands for assigning the value $a_l$ to the variable $u_l$ in instance $F$, where $a_l \in D_l$ and $l = 1, 2, \cdots, i$. $F(u_1/a_1, \cdots, u_l/a_l, \cdots, u_i/a_i)$ is false if and only if there is at least one constraint that has no compatible tuple of values)*

*3. Set $i \leftarrow i + 1$.*

4. If $i > n$, then $a_1, a_2, \cdots, a_n$ is a solution. Go to 7.

5. Set $u_i \leftarrow$ the first value in $D_i$ and go to 2.

6. If $u_i$ has more values, set $u_i \leftarrow$ the next value in $D_i$ and go to 2.

7. Set $i \leftarrow i - 1$. If $i > 0$, go to 6; otherwise stop.

Theoretical evaluation of constraint satisfaction algorithms is accomplished primarily by worst-case analysis[14]. However, a worst-case result often tells us relatively little about the behavior of algorithms in practice. Another way to evaluate the performance of an algorithm is to study the average time used by it on random problems. The time needed by the most basic form of backtracking on several different distributions of random problems has been studied in the previous papers[3],[7],[21],[22]. But these studies all used conjunctive normal form problems, where each variable had only two possible values. This paper focuses attention on random constraint satisfaction problems where each variable can have more values in its domain. Recently, Purdom[20] presented the first asymptotic analysis of the average speed of backtracking for solving random CSP. In Ref. [20], a random problem is formed by selecting with repetition $t$ random constraints. A random constraint is formed by selecting without repetition $k$ of $n$ variables, and each tuple of values of the $k$ variables are selected to be compatible with probability $p$. It is shown that the average number of nodes is polynomial when the ratio of constraints to variables is large, and it is exponential when the ratio is small. In fact, there are several ways of generating random CSP instances. In this paper, we propose a random CSP model that is as follows:

**Model GB**

Step 1. *We select with repetition $t$ random constraints. A random constraint is formed by selecting without repetition $k$ of $n$ variables.*

Step 2. *For each constraint we uniformly select without repetition $p \cdot d^k$ incompatible tuples of values, i.e., each constraint relation contains exactly $(1-p) \cdot d^k$ compatible tuples of values.*

Standard Model B[11],[24], commonly used for generating CSP instances, is actually a special case of Model GB with $k = 2$. That is to say, Model GB is a natural generalization of standard Model B to the $k$-ary case. We assume that $k \geq 2$ and all the variable domains contain the same number of values $d \geq 2$ in Model GB. Another condition is that the constraint tightness that determines how restrictive the constraints are satisfies the inequality $0 < p < 1/d^{k-1}$. We will show in the next section that this condition is sufficient for Model GB to avoid trivial asymptotic behaviour. A model suffering from *trivial asymptotic insolubility* means that the instances generated by this model are trivially unsatisfiable with probability tending to 1 as the number of variables approaches infinity. In such a case, it makes no sense to compute whether an instance is satisfiable since one knows that, asymptotically, no solution exists to the generated instances. A similar concept is *trivial asymptotic solubility*, which means that asymptotically, the instances are trivially satisfiable. Another thing worth noting is that the well-studied random $k$-SAT is a also special case of Model GB if we set $d$ to 2 and $p$ to $1/2^k$ respectively. It was found experimentally that the probability of an instance of random 3-SAT being satisfiable shifts with the ratio of clauses to variables, from being almost 1 with ratios below 4 to being almost 0 at ratios above 4.5[2]. The range of ratio over which this transition occurs becomes smaller as the number of variables increases. Another phenomenon is that the peak in difficulty occurs near the ratio where about half of the instances are satisfiable. The same pattern was also found for larger values of $k$. Because the instances generated in the transition region appear hardest to solve they are widely used in the experimental studies to evaluate the performance of algorithms and help us to design more efficient algorithms. Therefore, we can say that if a random CSP model suffers from trivial asymptotic behaviour which also means that no phase transition occurs, then this model will be asymptotically uninteresting for study. Since the phase transition phenomena were found in $k$-SAT and some other combinatorial problems[5], random CSP has also received great attention in recent years, both from an experimental and a theoretical point of view[1],[2],[8]~[13],[14]~[19],[23]~[26]. However, there is still some lack of studies about the probabilistic analysis of random CSP models.

This paper mainly analyzes the average complexity of backtracking on random constraint satisfaction problems. In section 1, we first give a brief introduction of CSP and then propose a random CSP model which is essentially a generalization of standard Model B. Section 2 presents a probabilistic analysis of Model GB. It is shown that Model GB will not suffer from trivial behaviour as the number of variables approaches infinity. In section 3, we give a detailed analysis obtain an asymptotic estimate of the average number of nodes in a search tree used by backtracking on Model GB. Section 4 investigates the behaviour of the average number of nodes. It is shown that the average number of nodes required for finding all solutions or proving that no solution exists grows exponentially with the number of variables. In addition, we further examine the behaviour of the average number of nodes as $r$ (the ratio of constraints to variables) varies. The results indicate that as $r$ increases random CSP instances get easier and easier to solve, and the base for the average number of nodes that is exponential in $n$ tends to 1 as $r$ approaches infinity.

## 2. Probabilistic analysis of Model GB

Recently, a theoretical result by Achlioptas *et al.*[1] shows that many models commonly used for generating CSP instances become trivially unsatisfiable as the number of variables increases. As we mentioned above, if a model suffers from trivial asymptotic insolubility, then it will be uninteresting for study. In this section we will prove that Model GB can avoid this problem. That is to say, when the ratio of clauses to variables is smaller than a certain value, the instances generated by Model GB are satisfiable with probability bounded from below by a positive constant as the number of variables approaches infinity. We prove this result by analyzing the behaviour of an algorithm for Model GB. In what follows, $C_1$ denote the set of constraints of arity 1. This algorithm, basically a natural extension of the Unit Clause heuristic for $k$-SAT introduced by Chao and Franco[4] is as follows:

*Algorithm: UC*

*Input: A random CSP instance*

*Output: "a solution exists" or "can not determine whether a solution exists"*

1. *Set $j \leftarrow 0$.*

2. *Repeat.*

3. *If $C_1 \neq \phi$, then choose, at random, a constraint $l$ from $C_1$ and assign a value to the variable $u_j$ in $l$ to make $l$ satisfied[1].*

4. *Else choose, at random, a variable $u_j$ not set yet and assign a value to it at random.*

5. *For each constraint containing $u_j$, check if the value assigned to $u_j$ in step 3 or step 4 occurs in the values assigned to $u_j$ in the incompatible tuples of values of this*

---

[1] Note that $p < 1/d^{k-1}$. The total number of incompatible tuples for a constraint is less than $d$ while each variable has $d$ possible values. So we can always find a value to satisfy $l$.

*constraint. If not, then remove this constraint.*

*6. Remove all the occurrences of $u_j$ from the constraints that contain it.*

*7. Set $j \leftarrow j+1$.*

*8. Until all the constraints are removed or an empty constraint is produced[2].*

*9. If all the constraints are removed Then Output ("a solution exists").*

*10. Else Output("can not determine whether a solution exists").*

*11. End.*

Note that Achlioptas *et al.* [1] have already introduced a UC algorithm adapted to CSP instances to prove that a CSP model proposed by them does not suffer from trivial asymptotic insolubility. From the procedure of the above algorithm it is not hard to see that if $C_1 = \phi$, the UC algorithm assigning a value to a variable in step 4 will not violate any constraint. But if $C_1 \neq \phi$, there is a possibility that an empty constraint will be produced, making the algorithm fail to find a solution. It should be noted that the concept of an empty constraint is very similar to that of an empty clause produced by two unit clauses in random $k$-SAT, which can help us to gain a better understanding of this algorithm. Chao and Franco analyzed the probabilistic performance of Unit Clause heuristic for $k$-SAT as a function of the ratio of clauses to variables. By use of their analysis, it is strightforward to obtain the following theorem:

**Theorem 1.** *If* $r < \dfrac{2d^{k-2}}{k(d-1)^{k-2}} \left( \dfrac{k-1}{k-2} \right)^{k-2}$, *then UC algorithm verifies that a solution exists for random CSP instances generated by Model GB with probability greater than $\varepsilon$ for some $\varepsilon > 0$ as the number of variables tends to infinity. For $k=2$, this condition amounts to $r < 1$.*

The basic idea behind the proof of this theorem, as given in [4], is as follows: after the algorithm has successfully assigned values to the first $j$ variables, there are $C_i(j)$ constraints of arity $i$ that are uniformly distributed among all possible constraints of arity $i$ on the unset variables. Moreover, if the average number of constraints of arity 1 into $C_1$ is less than 1 per step, then the number of constraints in $C_1$ will not, in probability, grow very large since at least one constraint is removed from $C_1$ whenever $C_1 \neq \phi$. In this case the probability that an empty constraint is produced is very small. Consequently, UC has a high probability of successfully finding a solution to a random instance. The condition in Theorem 1 is sufficient to guarantee that the average number of constraints of arity 1 into $C_1$ is less than 1 per step throughout the execution of the algorithm.

---

[2] For two constraints of arity 1 containing the same variable, if one constraint is satisfied by a value in step 3 but the other constraint can not be removed in step 5, then an empty constraint will be produced in step 6.

Theorem 1 establishes a region where asymptotically, a random CSP instance generated by Model GB is satisfiable with probability greater than a fixed positive constant. It means that Model GB does not suffer from trivial asymptotic insolubility whenever $p < 1/d^{k-1}$. Recall that standard Model B is a special case of Model GB with $k = 2$. Hence we can immediately arrive at a corollary[3] from Theorem 1 that Model B avoids trivial asymptotic insolubility whenever $p < 1/d$. This refutes a conjecture of Achlioptas *et al.* [1] at CP97[4] who proved that Model B suffers from trivial asymptotic insolubility whenever $p \geq 1/d$, and conjectured that Model B still suffers from trivial insolubility even when $p < 1/d$. Combining the result of Achlioptas *et al.* with Theorem 1, we can reach a conclusion that the condition $p < 1/d^{k-1}$ is not only sufficient but also necessary for Model GB with $k = 2$, i.e. Model B to avoid trivial asymptotic insolubility. However, this also leaves an open question, i.e. whether the condition $p < 1/d^{k-1}$ is still necessary for Model GB with $k \geq 3$.

After completing the above analysis, one may ask whether Model GB suffers from trivial asymptotic solubility. Fortunately, this problem is much simpler to deal with. First, the expected number of solutions $E(N)$ for model GB is given by

$$E(N) = d^n (1-p)^{rn}, \tag{1}$$

i.e. the number of possible assignments of $d$ values to $n$ variables, multiplied by the probability that a randomly-chosen assignment is compatible. Let $\Pr(Sat)$ denote the probability that a random CSP instance generated following Model GB is satisfiable. Note that $p > 0$, by the Markov inequality $\Pr(Sat) \leq E(N)$ we can then easily prove that

$$\lim_{n \to \infty} \Pr(Sat) = 0 \text{ when } r > r_{cr} = -\ln d / \ln(1-p). \tag{2}$$

---

[3] Ian Gent[11] also obtained this result.

[4] In the paper[1] submitted to *constraints*, the authors referred to the above result about Model B.

Combining Theorem 1 with equation (2), we find that Model GB does avoid trivial asymptotic behaviour. As mentioned in Section 1, one such example is the well-studied random $k$-SAT, which exhibits non-trivial behaviour as the number of variables tends to infinity.

## 3. The average number of nodes in a search tree

In this section we will first give an exact expression of the average number of nodes in a search tree, and then derive an asymptotic estimate of it through detailed asymptotic analysis. To calculate the average number of nodes, we first examine the probability that a random constraint on level $i$ has at least one compatible tuple of values (Root node is on level 0, and there are $i$ variables that have been assigned values on level $i$.). This probability is denoted by $g(i)$ in this paper.

If $i \leq k-1$, since each constraint contains $k$ variables, there must be a variable that has not been assigned values. Thus there are at least $d$ tuples of values for the constraint. Note that $p < 1/d^{k-1}$, so the total number of incompatible tuples for the constraint is less than $d$. Hence there must be a compatible tuple of values satisfying the constraint.

Thus we get

$$g(i) = 1. \tag{3}$$

If $i \geq k$, the probability that $k$ variables in a random constraint have all been assigned values is equal to $C_i^k / C_n^k$. In this case $g(i)$ is equal to $(1-p)$; otherwise $g(i)$ is equal to 1 (similar to the above analysis). Hence we have

$$g(i) = (1-p)\frac{\binom{i}{k}}{\binom{n}{k}} + (1-\frac{\binom{i}{k}}{\binom{n}{k}}) = 1 - p\frac{\binom{i}{k}}{\binom{n}{k}} = 1 - p\frac{i(i-1)...(i-k+1)}{n(n-1)...(n-k+1)}. \tag{4}$$

Combining equations (3) and (4) gives

$$g(i) = 1 - p\frac{i(i-1)...(i-k+1)}{n(n-1)...(n-k+1)}, \quad 0 \leq i \leq n-1. \tag{5}$$

Since each constraint is generated independently, the probability that all the $t = rn$ random constraints have at least one compatible tuple of values is equal to $[g(i)]^{rn}$. This is also the probability that node on level $i$ will extend to level $i+1$. Each extension of a node will result in $d$ additional nodes. There are $d^i$ possible nodes on level $i$. Consequently, the average number of nodes required for finding all solutions or proving that no solution exists for a random problem is

$$T_{av} = 1 + d \sum_{i=0}^{i=n-1} d^i [g(i)]^{rn}. \tag{6}$$

We now start to estimate $g(i)$ when $n$ approaches infinity. First, we have

$$p \frac{i(i-1)...(i-k+1)}{n(n-1)...(n-k+1)} = p \frac{\frac{i}{n}(\frac{i}{n}-\frac{1}{n})...(\frac{i}{n}-\frac{k-1}{n})}{(1-\frac{1}{n})...(1-\frac{k-1}{n})}. \tag{7}$$

Let $x = \frac{i}{n}$. It is obvious that $0 \le x \le 1$. We can easily prove that when $n$ is sufficiently large, the following inequality holds:

$$1 + \frac{j}{n} < \frac{1}{1-\frac{j}{n}} < 1 + \frac{j}{n} + 2\frac{j^2}{n^2}, \text{ where } j = 1, 2, \cdots, k-1. \tag{8}$$

By use of the above inequality and equation (7), we have

$$1 - px(x-\frac{1}{n})...(x-\frac{k-1}{n})(1+\frac{1}{n}+2\frac{1}{n^2})\cdots(1+\frac{k-1}{n}+2\frac{(k-1)^2}{n^2}) < g(i)$$

$$< 1 - px(x-\frac{1}{n})(x-\frac{2}{n})...(x-\frac{k-1}{n})(1+\frac{1}{n})...(1+\frac{k-1}{n}). \tag{9}$$

It should be noted that $k$, denoting the number of variables in a constraint, is a constant in the above equations. Rearranging the above inequality, we get

$$1 - px^k + \frac{k(k-1)p}{2n}(x^{k-1} - x^k) + \frac{D_2(x)}{n^2} + \frac{D_3(x)}{n^3} + \cdots + \frac{D_{3k-3}(x)}{n^{3k-3}} < g(i)$$

$$< 1 - px^k + \frac{k(k-1)p}{2n}(x^{k-1} - x^k) + \frac{E_2(x)}{n^2} + \frac{E_3(x)}{n^3} + \cdots + \frac{E_{2k-2}(x)}{n^{2k-2}}. \tag{10}$$

It is obvious that $D_j(x)$ and $E_l(x)$ are continuous functions, where $j = 1, 2, \cdots, 3k-3$ and $l = 1, 2, \cdots, 2k-2$. Let $D_{jm}$ and $E_{lM}$ stand for the minimum of $D_j(x)$ and the maximum of $E_l(x)$ on the interval $[0,1]$ respectively. Let $D_m = \min\{D_{1m}, D_{2m}, \cdots, D_{(3k-3)m}\}$ and $E_M = \max\{E_{1M}, E_{2M}, \cdots, E_{(2k-2)M}\}$. It is straightforward demonstrate that when $n$ is sufficiently large, the following inequality holds:

$$\frac{D_2(x)}{n^2} + \frac{D_3(x)}{n^3} + \cdots + \frac{D_{3k-3}(x)}{n^{3k-3}} > \frac{D_m}{n^2} + \frac{D_m}{n^3} + \cdots + \frac{D_m}{n^{3k-3}},$$

$$\frac{E_2(x)}{n^2} + \frac{E_3(x)}{n^3} + \cdots + \frac{E_{2k-2}(x)}{n^{2k-2}} < \frac{E_M}{n^2} + \frac{E_M}{n^3} + \cdots + \frac{E_M}{n^{2k-2}}. \quad (11)$$

By use of inequalities (10) and (11), we can easily prove that there exist two positive constants $C_0$ and $M_0$ such that

$$1 - px^k + \frac{k(k-1)p}{2n}(x^{k-1} - x^k) - \frac{C_0}{n^2} < g(i)$$

$$< 1 - px^k + \frac{k(k-1)p}{2n}(x^{k-1} - x^k) + \frac{C_0}{n^2} \text{ whenever } n > M_0. \quad (12)$$

Note that $i = n\frac{i}{n} = nx$, we get

$$T_{av} = 1 + d \sum_{i=0}^{i=n-1} d^{n\frac{i}{n}} [g(i)]^{rn} = 1 + \sum_{i=0}^{i=n-1} de^{n\frac{i}{n}\ln d} e^{rn \ln g(i)}. \quad (13)$$

It is easy to show that there exists a small positive constant $\varepsilon$ such that when $|z| < \varepsilon$, the following inequalities hold:

$$z - z^2 < \ln(1+z) < z \quad (14)$$

$$1 + z < e^z < 1 + z + z^2. \quad (15)$$

Let $\sigma(x) = \frac{k(k-1)p}{2}(x^{k-1} - x^k)$. With the help of inequalities (12) and (14), we obtain that when $n$ is sufficiently large, the following inequalities hold:

$$\ln(1 - px^k) + \frac{\sigma(x)}{n(1-px^k)} - \frac{C_0}{n^2(1-px^k)} - \left(\frac{\sigma(x)}{n(1-px^k)} - \frac{C_0}{n^2(1-px^k)}\right)^2 < \ln g(i)$$

$$< \ln(1-px^k) + \frac{\sigma(x)}{n(1-px^k)} + \frac{C_0}{n^2(1-px^k)} \tag{16}$$

Let $\varphi(x) = de^{\frac{\sigma(x)}{(1-px^k)}r}$ and $f(x) = x \ln d + r \ln(1-px^k)$. By inequalities (15), (16) and equation (13), we can easily prove that there exist two positive constants $C_1$ and $M_1$ such that

$$\varphi(x)e^{nf(x)}(1-\frac{C_1}{n}) < de^{n\frac{i}{n}\ln d} e^{rn \ln g(i)} < \varphi(x)e^{nf(x)}(1+\frac{C_1}{n}) \text{ whenever } n > M_1, \text{ where}$$

$$x = \frac{0}{n}, \frac{1}{n}, \cdots, \frac{n-1}{n}. \tag{17}$$

By use of equations (6), (11) and relation (17), we get

$$T_{av} = 1 + (\sum_{i=0}^{i=n-1} \varphi(\frac{i}{n}) e^{nf(\frac{i}{n})})(1+o(1)) \text{ when } n \to \infty. \tag{18}$$

When we derive the asymptotic estimate of equation (18), the following lemma 1 and lemma 2 in the appendix will be needed.

**Lemma 1.** *Given $r$, if $r > r_0 = \dfrac{(1-p)\ln d}{pk}$, then there is only one maximum point $0 < \zeta < 1$ of $f(x)$ on the interval $[0,1]$, and $\lim\limits_{r \to r_0^+} \zeta = 1$. If $r \leq r_0$, then the maximum point of $f(x)$ is at $x = 1$.*

*Proof.* Given $r$, to obtain the maximum of $f(x)$, we first analyze its derivatives:

$$f'(x) = \ln d - \frac{rpkx^{k-1}}{1-px^k}, \quad f''(x) = -rp\frac{k(k-1)x^{k-2} + pkx^{2k-2}}{(1-px^k)^2}, \tag{19}$$

$$f'(0) = \ln d, \quad f'(1) = (1-\frac{r}{r_0})\ln d. \tag{20}$$

If $r > r_0$, by equations (19), (20) we find that $f'(0) > 0$, $f'(1) < 0$, and $f''(x) < 0$ on the interval $(0,1)$. Then $f'(x)$ must be a strictly decreasing function. From the intermediate value theorem we know that there must be a unique point $0 < \zeta < 1$ such that $f'(\zeta) = 0$. Note that $f''(\zeta) < 0$, so $\zeta$ is the only one maximum point of $f(x)$. We now proceed to prove

$$\lim_{r \to r_0^+} \zeta = 1.$$

Let $x_0 = \left( \dfrac{\ln d}{\ln d + (r - r_0) pk} \right)^{\frac{1}{k-1}} \leq 1$. Substituting it into $f'(x)$ gives

$$f'(x_0) = \frac{p(x_0^{k-1} - x_0^{k}) \ln d}{1 - p x_0^{k}} \geq 0. \tag{21}$$

Since $f'(x)$ is a strictly decreasing function, the inequality $x_0 < \zeta < 1$ holds. From the expression of $x_0$, we obtain that $\lim_{r \to r_0^+} x_0 = 1$. Thus $\lim_{r \to r_0^+} \zeta = 1$ is proved.

If $r \leq r_0$, then $f'(1) \geq 0$. Note that $f'(x)$ is a strictly decreasing function, so $f'(x) > f'(1) \geq 0$ on the interval $[0,1)$. Thus $f(x)$ is a strictly increasing function on the interval $[0,1]$. Hence the maximum point of $f(x)$ is at $x = 1$.

Now we start to derive the asymptotic estimate of equation (18). Given $r$, let $F(r)$ denote the maximum of $f(x)$ on the interval $[0,1]$. By lemma 2 in the appendix we get

$$T_{av} = 1 + p(r) e^{nF(r)} (1 + o(1)) \text{ when } n \to \infty, \tag{22}$$

where $p(r) = \varphi(\zeta) \sqrt{\dfrac{2n\pi}{-f''(\zeta)}}$ when $r > r_0$; $p(r) = \dfrac{\varphi(1)}{2} \sqrt{\dfrac{2n\pi}{-f''(1)}}$ when $r = r_0$;

$p(r) = \dfrac{\varphi(1)}{d^{1-\frac{r}{r_0}} - 1}$ when $r < r_0$.

From equation (22) it is obvious that $p(r)$ is a polynomial function in $n$, and the behaviour of $T_{av}$ is mainly determined by the exponent. By analyzing the behaviour of $F(r)$, we can obtain the average case results for the backtracking algorithm on Model GB, which is the content of the next section. Moreover, it should be mentioned that there is a much simpler way to derive the exponent of $T_{av}$, as done in [20], [22]. But it will be very useful to estimate the average number of nodes more accurately in some cases such as in the experimental studies. So in this paper we presents a detailed analysis to derive an asymptotic estimate of the average number of nodes in a search tree for Model GB which is good to $1 + o(1)$.

## 4. Main results

**Theorem 2.** *Given $r$, the average number of nodes $T_{av}$ grows exponentially with $n$.*

*Proof.* By equation (22) we only need to prove $F(r) > 0$. From lemma 1, given $r > r_0$, there is only one maximum point $\zeta$ of $f(x)$. Thus we can define a function $\zeta(r)$ that varies with $r$. It is obvious that $\zeta(r)$ satisfies the following equation:

$$f'(\zeta(r)) = 0 \Rightarrow \ln d - \frac{rpk\zeta^{k-1}(r)}{1 - p\zeta^k(r)} = 0. \tag{23}$$

The proof of $F(r) > 0$ is divided into the following two cases:

**Case 1.** If $r > r_0$, then

$$F(r) = \zeta(r) \ln d + r \ln[1 - p\zeta^k(r)]. \tag{24}$$

Solving for $r$ from equation (24), in terms of $\zeta(r)$, and substituting it into equation (24), we get

$$F(r) = \frac{\ln d}{kp\zeta^{k-1}(r)} \{kp\zeta^k(r) + [1 - p\zeta^k(r)]\ln[1 - p\zeta^k(r)]\}. \tag{25}$$

Let $y = p\zeta^k(r)$, and $H(y) = ky + (1-y)\ln(1-y)$. Then

$$F(r) = \frac{\ln d}{kp\zeta^{k-1}(r)} H(y), \text{ where } 0 < y \le p. \tag{26}$$

$H(0) = 0$, $H'(y) = k - 1 - \ln(1-y) > 0$. So $H(y) > 0$. Hence $F(r) > 0$.

**Case 2.** If $r \le r_0$, then

$$F(r) = f(1) = \ln d + r \ln(1 - p). \tag{28}$$

It is obvious that $F(r)$ is a strictly decreasing function. Substituting $r_0$ into the above equation gives

$$F(r_0) = [1 - \frac{1-p}{kp} \ln(1 + \frac{p}{1-p})] \ln d. \tag{29}$$

It can be easily proved that $\frac{1-p}{p} \ln(1 + \frac{p}{1-p}) < 1$. Thus $F(r) \ge F(r_0) > 0$.

Theorem 2 shows that when we use the backtracking algorithm to solve Model GB, the average number of nodes required for finding all solutions or proving that no solution exists grows exponentially with $n$. Therefore, the average number of nodes used by the backtracking algorithm on the random CSP model is exponential.

**Theorem 3.** *Given $r_1$ and $r_2$, if $r_1 < r_2$, then* $\lim_{n \to \infty} \dfrac{T_{av}(r_2)}{T_{av}(r_1)} = 0$.

*Proof.* By equation (22) we only need to prove that $F(r)$ is a strictly decreasing function. The proof falls into the following two cases:

**Case 1.** If $r > r_0$, then

$$F'(r) = (\ln d - \frac{rpk\zeta^{k-1}(r)}{1 - p\zeta^k(r)})\zeta'(r) + \ln[1 - p\zeta^k(r)]. \tag{30}$$

Substituting (23) into (29) yields

$$F'(r) = \ln[1 - p\zeta^k(r)] < 0. \tag{31}$$

**Case 2.** If $r \leq r_0$, then

$$F(r) = f(1) = \ln d + r \ln(1 - p). \tag{32}$$

It is obvious that $F(r)$ is a strictly decreasing function. By lemma 1 we have $\lim_{r \to r_0^+} \zeta(r) = 1$. Hence $\lim_{r \to r_0^+} F(r) = \lim_{r \to r_0^+} f(\zeta(r)) = f(1)$, i.e. $F(r)$ is continuous at $r_0$. Combining the above two cases leads to lemma 2.

Note that $r$ is the ratio of constraints to variables, which determines how many constraints exist in a random CSP instance. Theorem 2 indicates that when $n$ is sufficiently large, search cost for instances with more constraints is much less than that for instances with fewer constraints. In other words, it gets easier and easier to solve the random CSP instances generated by Model GB as $r$ increases.

**Theorem 4.** As $r$ approaches infinity, the base for the average number of nodes which is exponential in $n$ tends to 1.

*Proof.* By equation (22) we only need to prove $\lim_{r \to \infty} F(r) = 0$. We first prove $\lim_{r \to +\infty} \zeta(r) = 0$.

Let $x_0 = \left(\dfrac{\ln d}{rkp}\right)^{\frac{1}{k-1}}$. When $r$ is sufficiently large, we have $0 < x_0 < 1$.

It is obvious that $f'(x_0) = \dfrac{-px_0^k \ln d}{1 - px_0^k} < 0$. Since $f'(x)$ is a strictly decreasing function, we obtain that $0 < \zeta(r) < x_0 = \left(\dfrac{\ln d}{rkp}\right)^{\frac{1}{k-1}}$. Hence $\lim\limits_{r \to +\infty} \zeta(r) = 0$.

$$\lim_{r \to \infty} F(r) = \ln d \cdot \lim_{r \to \infty} \zeta(r) + \dfrac{\ln d}{kp}[1 - \lim_{r \to \infty} p\zeta^k(r)] \lim_{r \to \infty} \dfrac{\ln[1 - p\zeta^k(r)]}{\zeta^{k-1}(r)}. \tag{33}$$

It can be easily proved that $\lim\limits_{y \to 0} \dfrac{\ln(1 - py^k)}{y^{k-1}} = 0$. Thus $\lim\limits_{r \to \infty} F(r) = 0$.

Theorem 3 shows that random CSP instances become easier and easier to solve as $r$ increases. Theorem 4 further proves that as $r$ goes to infinity, the base for the average number of nodes which is exponential in $n$ tends to 1. Therefore, when $r$ is sufficiently large, although the average number of nodes is still exponential, many random CSP instances will be very easy to solve.

## 5. Conclusions and future work

In this paper we proposed a random CSP model, called Model GB, which is a generalization of standard Model B. It is proved that Model GB exhibits non-trivial behaviour as the number of variables approaches infinity. An asymptotic analysis of the average number of nodes in a search tree used by the backtracking algorithm on Model GB was also presented. From Theorem 2 we know that the average number of nodes is exponential in the number of variables. So this model might be an interesting distribution for studying the nature of hard instances and evaluating the performance of CSP algorithms. We also investigated the behaviour of the average number of nodes as $r$ varies. Theorem 3 shows that random CSP instances become easier and easier to solve as $r$ increases. Theorem 4 further indicates that when $r$ is sufficiently large many random CSP instances will be very easy to solve. Note that random $k$-SAT is a special case of Model GB with $d = 2$ and $p = 1/2^k$. Thus we can immediately reach a conclusion that all the theorems in this paper hold for random $k$-SAT. As a result, Theorem 3 and Theorem 4 can help us to explain the contradiction between the experimental finding that random $k$-SAT with large $r$ is easy and the exponential average running time.

As mentioned in Section 1, phase transition behaviour not only is an important feature of random CSP but also has wide applications in the experimental studies. But in this paper our main focus was put on the average analysis of finding all solutions or proving that no solution exists, which can not shed any light on the peak in the average hardness of determining whether an instance is satisfiable. We suggest that future work should include this point which seems to be more complicated to analyze.

**Acknowledgement**

We would like to thank Yannis C. Stamatiou and two anonymous referees for their helpful comments and suggestions.

## Appendix

**Lemma 2.** *Given $r$, the following equation holds:*

$$\sum_{i=0}^{i=n-1} \varphi(\frac{i}{n}) e^{nf(\frac{i}{n})} = p(r)e^{nF(r)}(1+o(1)) \text{ when } n \to \infty,$$

where $\quad p(r) = \varphi(\zeta)\sqrt{\dfrac{2n\pi}{-f''(\zeta)}} \quad$ when $\quad r > r_0$; $\quad p(r) = \dfrac{\varphi(1)}{2}\sqrt{\dfrac{2n\pi}{-f''(1)}} \quad$ when $\quad r = r_0$;

$p(r) = \dfrac{\varphi(1)}{d^{1-\frac{r}{r_0}} - 1} \quad$ when $\quad r < r_0$.

*Proof.* Case 1: $r > r_0$.

From lemma 1 we know $f''(\zeta) < 0$. Let $\delta$ be a sufficiently small positive constant such that

$$f''(x) \leq -s < 0, \quad x \in [\zeta - \delta, \zeta + \delta].$$

Therefore, we have

$$\sum_{i=0}^{i=n-1} \varphi(\frac{i}{n}) e^{n(f(\frac{i}{n}) - f(\zeta))} = I_1 + I_2 + I_3 + I_4 + I_5, \tag{1}$$

where $I_1 = \sum_{i=0}^{i<n(\zeta-\delta)} Q(\frac{i}{n})$, $I_2 = \sum_{i \geq n(\zeta-\delta)}^{i<(n\zeta-n^{\frac{3}{5}})} Q(\frac{i}{n})$, $I_3 = \sum_{i \geq (n\zeta-n^{\frac{3}{5}})}^{i \leq (n\zeta+n^{\frac{3}{5}})} Q(\frac{i}{n})$, $I_4 = \sum_{i>(n\zeta+n^{\frac{3}{5}})}^{i \leq n(\zeta+\delta)} Q(\frac{i}{n})$,

$I_5 = \sum_{i>n(\zeta+\delta)}^{i \leq n-1} Q(\frac{i}{n})$, $Q(\frac{i}{n}) = \varphi(\frac{i}{n}) e^{n(f(\frac{i}{n}) - f(\zeta))}$.

Let $t_1 = \sup_{x \in [0, \zeta-\delta]} f(x) - f(\zeta)$. By lemma 1 we know that $t_1 < 0$. So

$$I_1 = \sum_{i=0}^{i<n(\zeta-\delta)} \varphi(\frac{i}{n}) e^{n(f(\frac{i}{n}) - f(\zeta))} = O(\sum_{i=0}^{i<n(\zeta-\delta)} \varphi(\frac{i}{n}) e^{nt_1}) = O(ne^{nt_1}). \tag{2}$$

Similarly, let $t_5 = \sup_{x \in [\zeta+\delta, 1]} f(x) - f(\zeta) < 0$. Then

$$I_5 = O(ne^{nt_5}). \tag{3}$$

An application of Taylor Theorem yields

$$f(x) - f(\zeta) = \frac{1}{2} f''(\eta)(x - \zeta)^2, \quad \eta = \zeta + \theta(x - \zeta), \quad 0 < \theta < 1.$$

Hence

$$I_2 = \sum_{i \geq n(\zeta-\delta)}^{i<(n\zeta-n^{\frac{3}{5}})} \varphi(\frac{i}{n}) e^{n(f(\frac{i}{n}) - f(\zeta))} = O(\sum_{i \geq n(\zeta-\delta)}^{i<(n\zeta-n^{\frac{3}{5}})} \varphi(\frac{i}{n}) e^{-n\frac{s}{2}(\frac{i}{2n} - \zeta)^2})$$

$$= O(\sum_{i \geq n(\zeta-\delta)}^{i<(n\zeta-n^{\frac{3}{5}})} \varphi(\frac{i}{n}) e^{-\frac{s}{2}n^{\frac{1}{5}}}) = O(ne^{-\frac{s}{2}n^{\frac{1}{5}}}). \tag{4}$$

Similarly, we get

$$I_4 = O(ne^{-\frac{s}{2}n^{\frac{1}{5}}}). \tag{5}$$

We now start to estimate $I_3$.

Let $i = (n\zeta + l)$ and $l = o(n)$. Expanding $f(x)$ in Taylor Series about $\zeta$, we obtain

$$f(\zeta + \frac{l}{n}) = f(\zeta) + f''(\zeta)\frac{l^2}{2n^2} + O(\frac{l^3}{n^3}), \quad \varphi(\zeta + \frac{l}{n}) = \varphi(\zeta)(1 + o(1)),$$

$$I_3 = \sum_{l \geq -n^{\frac{3}{5}}}^{l \leq n^{\frac{3}{5}}} \varphi(\zeta + \frac{l}{n})e^{n(f(\zeta+\frac{l}{n})-f(\zeta))} = \sum_{l \geq -n^{\frac{3}{5}}}^{l \leq n^{\frac{3}{5}}} \varphi(\zeta + \frac{l}{n})e^{f''(\zeta)\frac{l^2}{2n}+O(\frac{l^3}{n^2})}$$

$$= (1 + o(1))\varphi(\zeta) \sum_{l \geq -n^{\frac{3}{5}}}^{l \leq n^{\frac{3}{5}}} e^{f''(\zeta)\frac{l^2}{2n}}. \tag{6}$$

If $-n \leq l \leq -n^{\frac{3}{5}}$ or $n^{\frac{3}{5}} \leq l \leq n$, $e^{f''(\zeta)\frac{l^2}{2n}}$ is exponentially smaller than that of $l = 0$. So we can write equation (6) as

$$I_3 = (1 + o(1))\varphi(\zeta) \sum_{l=-n}^{l=n} e^{f''(\zeta)\frac{l^2}{2n}}. \tag{7}$$

Let $c = -f''(\zeta)$ and $H(x) = e^{-\frac{cx^2}{2n}}$. Applying Euler's summation formula (see [14], p.160 and [21]) to $\sum_{l=-n}^{l=n} H(l)$, we have

$$\sum_{l=-n}^{l=n} H(l) = (1 + o(1)) \int_{-\infty}^{\infty} H(x) dx = (1 + o(1))\sqrt{\frac{2n\pi}{c}}. \tag{8}$$

Combining the above results gives

$$\sum_{i=0}^{i=n-1} \varphi(\frac{i}{n})e^{n(f(\frac{i}{n})-f(\zeta))} = \varphi(\zeta)\sqrt{\frac{2n\pi}{-f''(\zeta)}}(1 + o(1)) + O(ne^{nt_1}) + O(ne^{nt_5}) + O(ne^{-\frac{s}{2}n^{\frac{1}{5}}})$$

$$= \varphi(\zeta)\sqrt{\frac{2n\pi}{-f''(\zeta)}}(1 + o(1)) \text{ when } n \to \infty. \tag{9}$$

Multiplying both sides of the above equation by $e^{nf(\zeta)}$, we obtain

$$\sum_{i=0}^{i=n-1} \varphi(\frac{i}{n}) e^{nf(\frac{i}{n})} = \varphi(\zeta) \sqrt{\frac{2n\pi}{-f''(\zeta)}} e^{nf(\zeta)}(1+o(1)) \text{ when } n \to \infty.$$

Case 2: $r = r_0$.

In this case, the asymptotic analysis is the same as that in Case 1, except that we only need to sum the terms from $l = -n$ to $l = 0$ in equation (7). Note that the maximum point of $f(x)$ is at $x = 1$ when $r = r_0$, we have

$$\sum_{i=0}^{i=n-1} \varphi(\frac{i}{n}) e^{nf(\frac{i}{n})} = \frac{\varphi(1)}{2} \sqrt{\frac{2n\pi}{-f''(1)}} e^{nf(1)}(1+o(1)) \text{ when } n \to \infty. \qquad (10)$$

Case 3: $r < r_0$.

Note that in this case the maximum point of $f(x)$ is also at $x = 1$. Let $i = n - l$ and $l = o(n)$. Expanding $f(x)$ in Taylor Series about $x = 1$, we obtain

$$f(1 - \frac{l}{n}) = f(1) - f'(1)\frac{l}{n} + O(\frac{l^2}{n^2}), \quad \varphi(1 - \frac{l}{n}) = \varphi(1)(1 + o(1)). \qquad (11)$$

By use of an asymptotic analysis similar to that in Case 1, we can easily show that only those terms near the maximum point of $f(x)$ have contributions to the asymptotic estimate. Hence we get

$$\sum_{i=0}^{i=n-1} \varphi(\frac{i}{n}) e^{nf(\frac{i}{n})} = \sum_{l=1}^{l=n} \varphi(1 - \frac{l}{n}) e^{nf(1-\frac{l}{n})} = (1+o(1))\varphi(1) e^{nf(1)} \sum_{l=1}^{l<n^{\frac{1}{3}}} e^{-f'(1)l}. \qquad (12)$$

From equation (17) we know that $f'(1) > 0$ when $r < r_0$. Hence $0 < e^{-f'(1)} < 1$ when $r < r_0$. Substituting the expression of $f'(1)$ into the above equation gives

$$\sum_{i=0}^{i=n-1} \varphi(\frac{i}{n}) e^{nf(\frac{i}{n})} = \frac{\varphi(1)}{d^{1-\frac{r}{r_0}} - 1} e^{nf(1)}(1+o(1)) \text{ when } n \to \infty. \qquad (13)$$

Hence lemma 2 is proved.